\documentclass[runningheads]{llncs}
\usepackage{graphicx}
\usepackage{amsmath}
\usepackage{adjustbox}
\usepackage{amsfonts}
\usepackage{dsfont}

\begin{document}
\title{Local Differential Privacy: a tutorial}
%
%
\author{Bj\"orn Bebensee}
%
%
\institute{Seoul National University \\ \email{bebensee@snu.ac.kr}}
\maketitle              
\begin{abstract}
In the past decade analysis of big data has proven to be extremely valuable in many contexts. Local Differential Privacy (LDP) is a state-of-the-art approach which allows statistical computations while protecting each individual user's privacy. Unlike Differential Privacy no trust in a central authority is necessary as noise is added to user inputs locally. In this paper we give an overview over different LDP algorithms for problems such as locally private heavy hitter identification and spatial data collection. Finally, we will give an outlook on open problems in LDP.

\end{abstract}
\section{Introduction}

With the advance of big data analytics and its value for businesses there has been an ever-growing interest in collecting statistics from user data to improve products and to gain valuable insights. However, valuable information is often highly sensitive to the users and can not be collected without giving strong privacy guarantees to users. In the past decade an approach called Differential Privacy ~\cite{dwork2006calibrating} which gives strong privacy guarantees for users has risen to popularity. They key idea behind differential privacy is that a user is given plausible deniability by adding random noise to their input. In the centralized Differential Privacy setting noise is added to the database. As the type of noise added is known statistical queries can still be computed by filtering out the noise while maintaining each user's individual privacy. However, this approach to Differential Privacy requires users to have trust the database maintainer to keep their privacy.

A stronger privacy guarantee for each individual users can be given in the local setting as there is no need to trust a centralized authority. Local Differential Privacy (LDP) was first formalized in~\cite{kasiviswanathan2011can} although an equivalent definition under the name ``amplification" had already been introduced by~\cite{evfimievski2003limiting}. In the local setting each user encodes and perturbs their own inputs before transmitting them to the untrusted server. This server can then compute statistical queries on the input data. The local settings poses its own challenges however. As each user perturbs their input individually the total variance is higher and in fact depends on the number of participants $n$~\cite{bassily2017practical}. LDP has been a quickly developing field with many new state-of-the-art approaches to important problems in LDP coming out in recent years~\cite{bassily2017practical,bun2018heavy,chen2016private,wang2017locally,wang2018locally}. Although the key idea behind LDP is relatively old~\cite{warner1965randomized} it has only recently risen to popularity and seen practical deployments by major technology organizations such as Apple~\cite{apple2017ldp}, who use it to collect usage statistics and find commonly used emojis, new words that are not part of the dictionary yet and to improve user behaviour, Google~\cite{erlingsson2014rappor}, who use it in Chrome to collect commonly chosen homepages, settings, and other web browsing behaviour, and Microsoft~\cite{ding2017collecting}, who use it for their collection of telemetry data.

\medskip

In this paper we give an introduction to Local Differential Privacy and present some of its applications. We then provide an overview of current state-of-the-art solutions to problems in Local Differential Privacy such as general frequency oracles, heavy hitter identification, itemset mining and locally private spatial data collection. Finally, we give a short outlook on the future of Local Differential Privacy and present some of its open problems.

\section{Local Differential Privacy}

In differential privacy the introduction of noise by randomization is crucial to ensure privacy through plausible deniability. We will first introduce the centralized notion of differential privacy called $\epsilon$-differential privacy. \cite{dwork2006differential} defines $\epsilon$-differential privacy ($\epsilon$-DP) as follows:
\begin{definition}
A randomized function $\mathcal{K}$ gives $\epsilon$-differential privacy if for all data sets $D_1$ and $D_2$ differing on at most one element, and all $S \subseteq Range(\mathcal{K}),$
$$
\frac{\text{Pr}[\mathcal{K}(D_1) \in S]}{\text{Pr}[\mathcal{K}(D_2) \in S]} \leq e^{\epsilon}
$$
\end{definition}

One major problem with this centralized notion of Differential Privacy is that users still have to trust a central authority, namely the database maintainer, to keep their privacy. In order to be able to give users stronger privacy guarantees the concept of Local Differential Privacy (LDP) was introduced. The to LDP definition equivalent notion of ``amplification" was first introduced by~\cite{evfimievski2003limiting}, and while the idea of local privacy was first formalized in~\cite{kasiviswanathan2011can} it rose to prominence through the work by Duchi et al.~\cite{duchi2013local}. LDP algorithms have been implemented and used in practice and we will introduce two such deployments by Apple and Google. We will use the definition of LDP given by~\cite{erlingsson2014rappor}.

\begin{definition}
We say that an algorithm $\pi$ satisfies $\epsilon$-Local Differential Privacy where $\epsilon > 0$ if and only if for any input $v$ and $v'$
$$
\forall y \in Range(\pi): \frac{Pr[\pi(v) = y]}{Pr[\pi(v') = y]} \leq e^\epsilon
$$
where $Range(\pi)$ denotes every possible output of the algorithm $\pi$.
\end{definition}

For $\epsilon$-LDP the \emph{privacy loss} is captured by $\epsilon$. For $\epsilon = 0$ perfect privacy is ensured as $exp(0) = 1$ while $\epsilon = \infty$ gives no privacy guarantee. It is important to note however that the choice of $\epsilon$ is crucial in practice as the increase in privacy risks is proportional to $exp(\epsilon)$. Thus a privacy loss of $exp(1)$ and $exp(50)$ have very different implications~\cite{tang2017privacy}.

\subsubsection{Randomized response} Although Local Differential Privacy has only recently been gaining traction and popularity, the ideas behind it are much older. The key idea of Local Differential Privacy is \emph{randomized response}, a surveying technique first introduced by Warner in~\cite{warner1965randomized}. This survey technique can be used to get more accurate statistics on topics where survey respondents would like to remain confidentiality. To answer the question the respondent first toss a coin in secret and answers truthfully only if the coin comes up heads. In case of tails the respondent tosses a second coin in secret and answers ``Yes" if the coin comes up tails and answers ``No" if the coin comes up heads. This preserves respondents' privacy by giving them strong plausible deniability while allowing for computation of accurate population statistics.

Given that respondents comply with the protocol, respondents will answer the question truthfully 75\% of the time. An accurate estimate of the true number of ``Yes" answers can therefore be computed by $2(X-0.25)$ where $X$ refers to the fraction of respondents who answered in the positive. The survey participants are given a privacy guarantee of $\epsilon = ln(\frac{0.75}{1-0.75}) = ln(3)$~\cite{erlingsson2014rappor}.

Although randomized response is a very basic LDP mechanism it is used in many more sophisticated LDP algorithms. One such implementation utilizing randomized response is \emph{RAPPOR} from Google~\cite{erlingsson2014rappor} which uses a variation of randomized response and combines it with bloom filters. We will describe this implementation in more detail in section \ref{sec:rappor}.

\subsection{Challenges in the local model}

As the local model does not use any centralized database and instead enables collection of statistics from a distributed set of inputs while maintaining privacy it has gained a lot of popularity in recent years and even been adopted by Apple~\cite{apple2017ldp} and Google~\cite{erlingsson2014rappor}. However the local model comes with its own challenges. Specifically it is more difficult to construct efficient LDP protocols and maintaining a low error-bound. Bassily et al.~\cite{bassily2017practical} give a good example of noise in the local model: when trying to estimate the number of HIV positives, it suffices to add Laplace noise of magnitude $O(1/\epsilon)$ in a centralized database which is independent of the number of $n$ participants. In the local model however a very high number of participants is needed because the lower-bound of $\Omega(\sqrt{n}/\epsilon)$ is dependent on $n$~\cite{chan2012optimal}. As~\cite{bassily2017practical} recognizes this makes low time, space, and communication complexity on the user side in a practical implementation invaluable.

\subsection{Practical implementations}

\subsubsection{RAPPOR} \label{sec:rappor}

One real world deployment of LDP has been made by Google in the form of RAPPOR~\cite{erlingsson2014rappor} in Google Chrome. The source code of its implementation has also been published online and can be adopted and implemented by others\footnote{https://github.com/google/rappor}. Google has been interested in using LDP to collect statistics of its Google Chrome browser to gain better insights in its usage without compromising the user's privacy. Erlingsson et al.~\cite{erlingsson2014rappor} present an algorithm that enables such a privacy-preserving collection of usage statistics.

Responses to the server can be assumed to be a bit vector of length $k$ where each bit represents some property. For collection of statistics on categorical properties such as whether a user belongs to a group or uses a certain one bit is needed for each category. Therefore a bit vector of length $k$ can provide information on $k$ different categorical properties. In order to collect numerical values the response bits can for instance be mapped to ranges of values which are then reported. However it is also possible to collect statistics on non-categorical properties by use of Bloom filters in combination with randomized response.

In particular to report a value $v$ it is added to the bloom filter $B$ of size $k$ using $h$ hash functions. Then, in order to guarantee the user's privacy, a variation of randomized response called permanent randomized response is applied first and then a second variation called instantaneous randomized response is applied second. Permanent randomized response only determines a single bit vector $B'$ for each value $v$ which is then \emph{memoized} and reused every time in place of the real answer. This is done to ensure that no inference of the user's real answer is possible even if a value is reported multiple times. To compute the permanent randomized response given the user's longitudinal privacy guarantee parameter $f$, for each bit $i$ in $B$ a reporting value $B'_i$ is determined:
$$
B'_i =
\begin{cases}
1, & \text{with probability } \frac{1}{2}f \\
0, & \text{with probability } \frac{1}{2}f \\
B_i, & \text{with probability } 1-f
\end{cases}
$$
Each time the value $v$ is reported the bit vector $B'$ is reused to compute the actual response bit vector $S$ of length $k$ using instantaneous randomized response. Instantaneous randomized response works by perturbing each bit of $B'$ with randomized response parameters $p, q$. First a bit vector $S$ is initialized and all values are set to 0. Then each bit $i$ of S is set with probabilities
$$
P(S_i=1) =
\begin{cases}
q, & \text{if } B'_i = 1\\
p, & \text{if } B'_i = 0
\end{cases}
$$
Finally the vector $S$ is sent to the server.

\medskip

Additionally Erlingsson et al. present a few modifications of the RAPPOR algorithm. \emph{One-time RAPPOR} is a modification that does not require protection against longitudinal attacks, i.e. inference attacks on observations of multiple transmissions, as it is only a one-time report of a value. In this case the instantaneous randomized response step can be skipped and the result of direct randomization $B'$ transmitted instead. \emph{Basic RAPPOR} is a variation which is equivalent to the previously mentioned collection of statistics on categorical properties. Given a well-defined and small set of strings each of the strings can be mapped directly to one of the bits in the bit vector instead of using a bloom filter. \emph{Basic one-time RAPPOR} is the combination of the two variations. Permanent randomized response satisfies $\epsilon_1$-LDP where
$$\epsilon_1 = 2h \ ln \bigg(\frac{1-\frac{1}{2}f}{\frac{1}{2}f}\bigg)$$

As the probability of observing a 1 in the instantaneous randomized response step is a function of both $p$, $q$ and $f$ we calculate probabilities of observing a 1 given that the underlying Bloom filter bit was set or was not set as follows:
\begin{equation} \label{eq:q}
q' = P(S_i = 1|b_i = 1) = \frac{1}{2}f(p+q)+(1-f)q
\end{equation}
\begin{equation} \label{eq:p}
p' = P(S_i = 1|b_i = 0) = \frac{1}{2}f(p+1)+(1-f)p
\end{equation} 
Then, given (\ref{eq:q}) and (\ref{eq:p}), instantaneous randomized response satisfies $\epsilon_2$-LDP with
$$\epsilon_2 = h \ ln\bigg(\frac{q'(1-p')}{p'(1-q')}\bigg)$$

\subsubsection{Apple's implementation of LDP}

Another real-world deployment of LDP has been made by Apple~\cite{apple2017ldp} who use it to collect usage statistics to better understand user behavior and improve the user experience. Apple has been most interested in the problem of frequency estimation. One such example is the estimation of frequently used emojis while maintaining users' privacy by use of LDP. To give its users $\epsilon$-LDP guarantees, Apple uses \emph{count-mean sketch} (CMS), which is a variation of \emph{count-min sketch}~\cite{cormode2005improved}, a probabilistic sublinear space data structure that allows for efficient operations on data streams.

\bigskip

On the server-side this implementation uses a sketch matrix $M$ of dimensions $k \times m$ with $k$ hash functions. The client then maps domain elements $d \in \mathcal{D}$ to size $m$ allowing for a reasonable transmission size that does not impact the user. Similarly to the count operation in \emph{count-min sketch}, a frequency estimate for a domain element $d \in \mathcal{D}$ can be computed by averaging the counts corresponding to the according $k$ hash functions in $M$. Instead of taking the minimum count in $M$ like in \emph{count-min sketch}, the average count is computed in order to provide better accuracy for randomly perturbed sketch matrices.

\bigskip

On the client-side, in order to transmit a data entry $d \in \mathcal{D}$ while maintaining $\epsilon$-LDP for a given privacy parameter $\epsilon > 0$, a hash functions $h$ out of the available $k$ hash functions is chosen uniformly at random. Then, for an encoding vector $v \in \{-1, 1\}^m$ the entry at position $h(d)$ is set to 1 and every other entry is set to -1. At least, much like in randomized response, each bit of the encoding vector $v$ is flipped with probability $\frac{1}{1+e^{\epsilon/2}}$. This vector is then transmitted to the server where the sketch matrix $M$ is updated accordingly.

\bigskip

Apple also presents another variation of CMS in~\cite{apple2017ldp} called \emph{Hadamard count-mean sketch} (HCMS) which makes use of the Hadamard transform to achieve a similar variance to CMS while only transmitting a single bit to the server. However both CMS and HCMS assume that there is some known dictionary of domain elements which the server can query to find counts for all its data entries and to create a histogram. Apple presents an algorithm called Sequence Fragment Puzzle (SFP) which allows for calculation of histograms on unknown dictionaries. One example application of this is the discovery of new popular words that are not yet included in a phone's dictionary while maintaining $\epsilon$-LDP.

\bigskip

There has been some criticism of the practical deployment of LDP by Apple however. Although the algorithms presented can ensure users' privacy the choice of the privacy loss parameter $\epsilon$ is important. Tang et al.~\cite{tang2017privacy} criticize the lack of transparency in the choice of privacy loss permitted by the system. Upon closer examination of Apple's actual implementation they found that although a privacy loss of $\epsilon=1$ or $\epsilon=2$ was guaranteed for single transmissions, Apple allowed up to 16 such transmissions per day meaning the upper bound for privacy loss is as high as $\epsilon = 16$ per day. This shows that although first steps to adopt $\epsilon$-LDP in practice have been taken, not all implementations provide equal privacy guarantees or even transparency of what the privacy guarantees given are.

\subsection{Frequency Oracles}

One core problem of LDP is locally private frequency estimation. Given a domain $\mathcal{D}$ a \emph{frequency oracle} (FO) is a protocol which estimates the frequency of an element $d \in \mathcal{D}$. A basic FO protocol has first been proposed in~\cite{warner1965randomized}.

Wang et al.~\cite{wang2017locally} introduce an abstract framework which several other FO protocols can be placed in. They divide FO protocols into three main steps: for each question, the user encodes their answer, perturbs the encoded value and sends it to the aggregator who obtains a frequency estimate for each answer of the question by decoding the reported values. The generalized framework provides several improvements in accuracy over previous algorithms such as the basic RAPPOR protocol~\cite{erlingsson2014rappor}. First Wang et al. define pure LDP protocols given PE, the composition of the encoding and perturbation algorithms, and Support, which maps each output $y$ to a set of inputs that support that are supported by the output value $y$, as follows:

\begin{definition}
A protocol given by {\normalfont PE} and {\normalfont Support} is pure if and only if there exist two probability values $p^* > q^*$ such that for all $v_1$,
\begin{align*}
& \text{\normalfont Pr}[\text{\normalfont PE}(v_1) \in \{ y \ | \ v_1 \in \text{\normalfont Support}(y) \}] = p^* \\
\forall_{v_2 \neq v_1} & \text{\normalfont Pr}[\text{\normalfont PE}(v_2) \in \{ y \ | \ v_1 \in \text{\normalfont Support}(y) \}] = q^*
\end{align*}
\end{definition}

We call $\{ y \ | \ v_1 \in \text{\normalfont Support}(y) \}$ the support set of $v_1$. Then, $p^*$ represents the probability that any value $v_1$ is mapped to its own support set and $q^*$ denotes the probability that any other value is mapped to the support set of $v_1$. In order to satisfy $\epsilon$-LDP we must have $q^* > 0$ as it must be possible for a different value to be mapped to the support set of $v_1$ and specifically it is required that $\frac{p^*}{q^*} \leq e^{\epsilon}$. Pure LDP requires that $p^*$ be the same for all values and $q^*$ be the same for all pairs of values.

In a pure LDP setting the true frequency $c(i)$ of a value $i$ can be estimated by counting reported outputs that support $i$. However, because of the perturbation step we expect to see at least $n \cdot q^*$ and at most $n \cdot p^*$ outputs that support $i$. This unbiased estimate is given by
\begin{equation} \label{aggregator}
\hat{c}(i) = \frac{\sum_j \mathds{1}_{\text{Support}(y^j)}(i)-nq^*}{p^* - q^*}
\end{equation}
where $y^j$ is the value submitted by user $j$.

Next Wang et al. introduce their framework of pure LDP FO protocols which generalize many previously proposed protocols. They introduce four distinct types of protocols that utilize different encoding techniques. For two protocols they propose their own variants with optimal parameters (Optimized Unary Encoding and Optimized Local Hashing). As all of the protocols are pure LDP protocols the frequency of any value $d \in \mathcal{D}$ can be estimated by (\ref{aggregator}).

\subsubsection{Direct Encoding (DE)} Direct Encoding is the generalization of the basic FO protocol by Warner~\cite{warner1965randomized} with $\text{Encode}(v) = v$. The values are then perturbed with
$$
\text{Pr}[\text{Perturb}(x) = i] =
\begin{cases}
p = \frac{e^{\epsilon}}{e^{\epsilon}+|\mathcal{D}|-1}, & \text{if } i=x\\
q = \frac{1-p}{|\mathcal{D}|-1} = \frac{1}{e^{\epsilon}+|\mathcal{D}|-1}, & \text{if } i \neq x
\end{cases}
$$
The support function is $\text{Support}_{\text{DE}}(i) = \{ i \}$. The aggregator can estimate the frequency of any value $d \in \mathcal{D}$ using (\ref{aggregator}). However, as the variance is linear in $|\mathcal{D}|$ the accuracy suffers for bigger domains.

\subsubsection{Histogram Encoding (HE)} Histogram Encoding uses a histogram of length $|\mathcal{D}|$ to encode each input $d \in \mathcal{D}$. Specifically, for a value $v$ its encoding is a vector of length $|\mathcal{D}|$ in which each entry is 0.0 and only the $v$-th component equals 1.0. This vector is then perturbed using the Laplace distribution such that $B'[i] = B[i] + \text{Lap}(\beta)$ with $\text{Pr}[\text{Lap}(\beta) = x] = \frac{1}{2 \beta} e^{-|x|/\beta}$. For aggregation, Wang et al. present two techniques: Summation with Histogram Encoding (SHE) which sums up the reported noisy histograms from all users and is not a pure LDP protocol, and Thresholding with Histogram Encoding (THE) which is pure and interprets each noisy count above a threshold $\theta$ as a 1 and each count below $\theta$ as a 0. Its support function is given by $\text{Support}_{\text{THE}}(B) = \{ v \ | \ B[v] > \theta \}$. As noise of large magnitude does not sum up when using a threshold the variance of THE is lower than that of SHE.

\subsubsection{Unary Encoding (UE)} In Unary Encoding a value is encoded as a bit vector similarly to the approach used in the basic RAPPOR protocol~\cite{erlingsson2014rappor}. Specifically a value $v$ is encoded as a bit vector where only the $v$-th position equals 1 and all other positions equal 0. Given probabilities $p,q$ the perturbed output $B'$ is computed as follows:
$$
\text{Pr}[B'[i] = 1] = 
\begin{cases}
p, & \text{if } B[i] = 1\\
q, & \text{if } B[i] = 0
\end{cases}
$$
Depending on the choice of $p$ and $q$ Wang et al. differentiate between Symmetric Unary Encoding (SUE) which uses $p + q = 1$ and is equivalent to basic RAPPOR~\cite{erlingsson2014rappor} and their own variant Optimized Unary Encoding (OUE) with $p = \frac{1}{2}$ and $q = \frac{1}{e^{\epsilon}+1}$ which are optimal parameters that minimize the error. UE's support function is $\text{Support}_{\text{UE}}(B) = \{ i \ | \ B[i] = 1 \}$.

\subsubsection{Binary Local Hashing (BLH)}

They key idea behind Binary Local Hashing is that communication cost can be lowered by hashing the input values to a domain of size $k < |\mathcal{D}|$. BLH is logically equivalent to the random matrix-base protocol proposed by Bassily et al.~\cite{bassily2015local}. Consider a universal family of hash functions $\mathbb{H}$, such that each hash function $h \in \mathbb{H}$ hashes an input $d \in \mathcal{D}$ into one bit, and such that the family's universal property is given by
$$
\forall x,y \in \mathcal{D}, x \neq y: \text{Pr}_{H \in \mathbb{H}}[H(x)=H(y)] \leq \frac{1}{2}.
$$
A user's input $v$ can now be encoded as $\text{Encode}(v) = \langle H, b \rangle$ with $H \in \mathbb{H}$ chosen uniformly at random and $b = H(v)$. Given $\epsilon$ and the input bit $b$ the perturbed input $\langle H,b' \rangle$ is given by
$$
\text{Pr}[b' = 1] =
\begin{cases}
p = \frac{e^{\epsilon}}{e^{\epsilon}+1}, & \text{if } b = 1\\
q = \frac{1}{e^{\epsilon}+1}, & \text{if } b = 0
\end{cases}
$$
Its support function is $\text{Support}_{\text{BLH}}(\langle H,b \rangle) = \{ v \ | \ H(v) = b \}$. It is important to note that each reported value $\langle H,b \rangle$ supports half of the input values $d \in \mathcal{D}$ as the output is only a single bit of information.

\subsubsection{Optimal Local Hashing (OLH)}

As Wang et al. observe that information is lost in BLH as the output is only a single bit they generalize BLH and propose Optimal Local Hashing (OLH) which instead hashes each input value into a value in [g] where $g \geq 2$. The choice of $g$ is crucial as a larger value allows for more information to be preserved in the encoding step but leads to more information being lost in the random response step. They analytically determine the optimal parameter to be $g = e^{\epsilon} + 1$.

Given a universal family of hash functions $\mathbb{H}$ such that for every $H \in \mathbb{H}$ any input $d \in \mathcal{D}$ is mapped to a value in $[g]$. Then, a user input $v$ is again encoded as $\text{Encode}(v) = \langle H,b \rangle$ with $b = H(v)$ for $H \in \mathbb{H}$ chosen uniformly at random. The perturbed output is $\langle H,b' \rangle$ where
$$
\forall_{i \in [g]} \text{Pr}[b' = i] = 
\begin{cases}
p = \frac{e^{\epsilon}}{e^{\epsilon} + g - 1}, & \text{if } b = i\\
q = \frac{1}{e^{\epsilon} + g - 1}, & \text{if } b \neq i
\end{cases}
$$
Its support function is given by $\text{Support}_{\text{OLH}}(\langle H,b' \rangle) = \{ i \ | \ H(i) = b' \}$.

\bigskip

\noindent Wang et al. observe that OLH and OUE have the exact same variance and values for $p^*$ and $q^*$ although they use different approaches to encoding. OLH has a lower communication cost of $O(\text{log } |\mathcal{D}|)$ compared to OUE's cost of $O(|\mathcal{D}|)$. However, as these two protocols use different optimized encoding approaches but yield the exact same variance Wang et al. suggest that it is possible that these protocols may be optimal for large domains $\mathcal{D}$.

Finally, they analyzed which LDP protocol is best depending on the size of the domain $\mathcal{D}$. They came to the conclusion that DE is best for small domains with $|\mathcal{D}| < 3 e^{\epsilon} + 2$, and that OUE and OLH are better for domains $|\mathcal{D}| > 3 e^{\epsilon} + 2$. However, as OUE has communication cost $\Theta (|\mathcal{D}|)$ we should use OLH for domains where the communication cost becomes too big as it offers the same accuracy at lower communication cost of $O(\text{log } |\mathcal{D}|)$.

\subsection{Heavy Hitter Identification}

In the heavy hitter identification problem in the local model we wish to estimate the frequency of common domain elements (\emph{heavy hitters}) among a set of users. For small domains this can be done by using a simple frequency oracle protocol and estimating the frequency of all domain elements. This approach is computationally infeasible for larger domains however. The problem of heavy hitter identification is very well studied \cite{bassily2017practical,bassily2015local,bun2018heavy,hsu2012distributed,mishra2006privacy,qin2016heavy}. As the error bound for efficient heavy hitter protocols presented by Mishra et al.~\cite{mishra2006privacy} and Hsu et al.~\cite{hsu2012distributed} is higher, we will focus on more recent work by Bassily and Smith~\cite{bassily2015local} Bassily et al.~\cite{bassily2017practical} and Bun et al.~\cite{bun2018heavy}.

\bigskip

Formally we consider a set of $n$ users each holding an input $x_i \in \mathcal{D}$. Then $S$ is a ``distributed database" with $S = (x_1,\ldots,x_n)$ consisting of all users' inputs. Bun et al.~\cite{bun2018heavy} define a domain element $x \in \mathcal{D}$ as $\Delta$-heavy if its multiplicity in S is at least $\Delta$. This condition is satisfied if there are at least $\Delta$ users who hold the input $x$. For $\Delta$ as small as possible we want to find all $\Delta$-heavy elements (i.e. \emph{heavy hitters}). As all domain elements with multiplicities smaller than $\Delta$ are not excluded, the parameter $\Delta$ is also referred to as the protocol's \emph{error}.

An efficient protocol for heavy hitter identification has been presented by Bassily and Smith~\cite{bassily2015local}. First Bassily and Smith construct a protocol for computation of succinct histograms (S-Hist) that satisfies $\epsilon$-LDP. A succinct histogram is a data structure which provides a list of heavy hitters $(v_1, \ldots, v_n)$ and their estimated frequencies $\hat{f}(v_i) : i \in [n]$. The error of a succinct histogram is given by the Chebyshev distance $\ell_{\infty}$ between the estimated and actual frequency as $\text{max}_{v \in \mathcal{D}} |\hat{f}(v) - f(v)|$. Then, in order to keep communication cost low Bassily and Smith present a transformation that can transform any $(\epsilon,0)$-DP local protocol into a 1-bit protocol given that it uses a public coin model where both user and server have access to a common random string. Such a protocol only requires each user to send a single bit to the server for the statistical query to be computed. Using this transformation they finally transform the succinct histogram protocol into a 1-bit protocol.

Bassily et al. revisit the problem in~\cite{bassily2017practical} and present new efficient heavy hitter algorithms TreeHist and Bitstogram which achieve near-optimal worst-case error while improving server time complexity to $\tilde{O}(n)$ and user time complexity to $\tilde{O}(1)$. We will give a short overview over the TreeHist and Bitstogram algorithms below.

\subsubsection{TreeHist} For TreeHist a binary prefix tree whose leaves correspond to domain elements is constructed. The algorithm then scans the levels of the tree starting at the top level and pruning all nodes and their children which cannot be prefixes of a heavy hitters by making queries to a given frequency oracle for that prefix. Finally, when the algorithm reaches the bottom level of the tree, the heavy hitters and a more accurate estimate of their frequencies can be determined using the frequency oracle. The number of surviving nodes does not exceed $O \big( \sqrt{n/(\text{log}(d) \cdot \text{log} (n))} \big)$ with high probability and therefore at most $O \big( \sqrt{n \text{ log}(d) / \text{log}(n)} \big)$ nodes are queried with the frequency oracle.

\subsubsection{Bitstogram} In the Bitstogram protocol every user $j \in [n]$ has access to a matrix $Z \in \{ -1, 1 \}^{d \times n}$ chosen uniformly at random for a domain $\mathcal{D}$ of size $|\mathcal{D}| = d$. Every user $j$ can then send a randomized response of their input $v_j$ using the corresponding bit $Z[v_j,j]$ from the public matrix $Z$: they send $y_j = Z[v_j,j]$ with probability $\frac{1}{2} + \frac{\epsilon}{2}$ and $y_j = -Z[v_j,j]$ with probability $\frac{1}{2} - \frac{\epsilon}{2}$. A frequency oracle that estimates the frequency $a(v)$ of a domain element $v \in \mathcal{D}$ can then be constructed as follows:
$$
a(v) = \frac{1}{\epsilon} \cdot \sum_{j \in [n]} y_j \cdot Z[v,j]
$$
To identify heavy hitters the input domain is hashed into a domain of size $T = \sqrt{n}$ where $n$ is the number of users. Given a database $S_\ell = (h(v_j), v_j[\ell])_{j \in [n])}$ for a random hash function $h: \mathcal{D} \rightarrow [T]$ and an input $v_j \in \mathcal{D}$ where $v_j[\ell]$ is the $\ell$-th bit of $v_j$, the entry $(h(x),x[i])$ will have a significantly higher count than $(h(x),\neg x[i])$ for a heavy hitter $x$. This then allows for recovery of all bits of $x$ using the frequency oracle.

Most recent research on frequency oracles and heavy hitter algorithms has focused on minimizing the error. However as Bassily and Smith~\cite{bassily2015local} have shown the lower bound for the worst-case error of these tasks in LDP is at least $\Omega(\frac{1}{\epsilon} \sqrt{n \cdot log |X|})$. Although the above presented previous work by Bassily and Smith~\cite{bassily2015local} and Bassily et al.~\cite{bassily2017practical} was efficient and had a near-optimal worst-case error, it had sub-optimal dependency of the error on the failure probability $\beta$ (see table \ref{table_heavy_hitters})~\cite{bun2018heavy}.

Bun et al.~\cite{bun2018heavy} present PrivateExpanderSketch which is a heavy hitter algorithm with optimal theoretical performance. This heavy hitter algorithm is based on the frequency oracle Hashtogram in~\cite{bassily2017practical}. An overview over the performance of different heavy hitter algorithms is given in table~\ref{table_heavy_hitters}.

\begin{table}
\caption{Table from \cite{bun2018heavy} comparing different heavy hitter algorithms \cite{bassily2017practical,bassily2015local,bun2018heavy}. $\tilde{O}$ notation is used to hide logarithmic factors. Parameters are the number of users $n$, size of the domain $|X|$, privacy parameter $\epsilon$, and the failure probability $\beta$.} \label{table_heavy_hitters}
\adjustbox{width=\textwidth,center}{
\bgroup
\setlength{\tabcolsep}{0.5em}
\renewcommand\arraystretch{2}
\begin{tabular}{|c||c|c|c|}
    \hline
    Performance metric & Bun et al.~\cite{bun2018heavy} & Bassily et al.~\cite{bassily2017practical} & Bassily and Smith.~\cite{bassily2015local}\\
    \hline
    \hline
    Server Time & $\tilde{O}(n)$ & $\tilde{O}(n)$ & $\tilde{O}(n^{2.5})$ \\
    \hline
    User Time & $\tilde{O}(1)$ & $\tilde{O}(1)$ & $\tilde{O}(n^{1.5})$ \\
    \hline
    Server Memory & $\tilde{O}(\sqrt{n})$ & $\tilde{O}(\sqrt{n})$ & $\tilde{O}(n^2)$ \\
    \hline
    User Memory & $\tilde{O}(1)$ & $\tilde{O}(1)$ & $\tilde{O}(n^{1.5})$ \\
    \hline
    Communication/user & $\tilde{O}(1)$ & $\tilde{O}(1)$ & $\tilde{O}(1)$ \\
    \hline
    Public randomness/user & $\tilde{O}(1)$ & $\tilde{O}(1)$ & $\tilde{O}(n^{1.5})$ \\
    \hline
    Worst-case error & $O\bigg( \frac{1}{\epsilon} \cdot \sqrt{n \ \text{log} \big(\frac{|X|}{\beta}\big)} \bigg)$ & $O\bigg( \frac{1}{\epsilon} \cdot \sqrt{n \ \text{log} \big(\frac{|X|}{\beta}\big) \text{log} \big(\frac{1}{\beta}\big)} \bigg)$ & $O\bigg( \frac{\text{log}^{1.5}(\frac{1}{\beta})}{\epsilon} \cdot \sqrt{n \ \text{log} |X|} \bigg)$ \\
    \hline
\end{tabular}
\egroup
}
\end{table}

\subsection{Itemset mining} The itemset mining problem deals with the collection of statistics on set-valued inputs rather than single-valued inputs in a locally differential private setting. An example for this problem of frequency estimation on set-valued inputs is given by Thakurta et al.~\cite{thakurta2017emoji}. Here, Apple wants to estimate the frequency of emojis typed while the user submits a set of emojis that they typed during a given time period. As frequency oracle and heavy hitter protocols function by filtering out the added noise through a big enough population size an application of them to the itemset setting is not possible as transactions may appear only infrequently while their items and itemsets might appear frequently. Consider for example transactions $\{a,b,z\}$, $\{a,c,d,z\}$, $\{a,e,f,z\}$. Then we have frequent itemsets $\{a\}$, $\{z\}$ and $\{a,z\}$ that all occurred three times even though each transaction only occurred once.

Most work on this problem is very recent. A first approach to solving heavy hitter estimation over set-valued data was given by Qin et al.~\cite{qin2016heavy} who present a mechanism called LDPMiner. This mechanism works in a two-phase framework: in the first phase the items are filtered and potential heavy hitters are identified, in the second phase the frequency estimates are refined. $\epsilon$-LDP is guaranteed by splitting the privacy budget $\epsilon$ into $\epsilon_1$ and $\epsilon_2$ which are then allocated to the two phases. Phase I then works the same way previous heavy hitter algorithms did: each user reports their set of inputs under $\epsilon_1$-LDP and a list of $k_{max}$ items with the highest frequency is estimated. To achieve higher accuracy the data collector then broadcasts this set of $k_{max}$ items to all users. In Phase II each user then reports their inputs with $\epsilon_2$-LDP -- however this time the domain is reduced to the $k_{max}$ candidate items and therefore allowing for computation of a higher accuracy estimate.

Qin et al. use a sampling randomizer algorithm to deal with the itemset setting. Here, instead of reporting their entire itemset, each user randomly samples an item which they subsequently report. Assuming that every user would normally submit a set of $\ell$ items, the frequency oracle should multiply the estimated frequency by $\ell$ to achieve an unbiased estimate. However, it is important that users with less than $\ell$ items pad their itemset before sampling while users with more than $\ell$ items simply generate a new set of exactly $\ell$ items by random sample without replacement. Without padding the probability of any item being chosen during the sampling process is hard to assess which makes it hard to compute an unbiased frequency estimate.

Based on this sampling randomizer algorithm they present sampling succinct histogram (sampling SH), which is a variant of succinct histogram (S-Hist) by Bassily et al.~\cite{bassily2015local}, as well as a variant of RAPPOR~\cite{erlingsson2014rappor} called sampling RAPPOR. These sampling-randomizer-based methods can then be used in Phase I and II of the LDPMiner mechanism.

Work by Wang et al.~\cite{wang2018locally} proposes a new Set-Value Item Mining (SVIM) protocol to find frequent items in the set-valued setting which provides significantly higher accuracy than LDPMiner as well as the Set-Value itemSet Mining (SVSM) protocol which can be used to identify frequent itemsets rather than single items. Previous to this work, identifying frequent itemsets had still been an open problem as LDPMiner focused on identifying only frequent items (i.e. singleton itemsets) rather than itemsets.

Wang et al. investigate padding-and-sample-based frequency oracles (PSFO) in regards to the effect of privacy amplification which has previously been studied in the standard DP setting~\cite{li2012sampling}. Consider the sampling randomizer algorithm from~\cite{qin2016heavy}. Since the sampling step randomly selects an item, it provides an amplification effect in terms of privacy. As each item is selected with a probability $\beta = \frac{1}{\ell}$ the frequency oracle can be invoked with an $\epsilon' = \text{ln} (\ell \cdot ( e^{\ell} - 1) + 1) \geq \epsilon)$ while still guaranteeing $\epsilon$-LDP.

They found that whether such a privacy amplification effect applies depends on the FO protocol used. Generalized Random Response and Optimized Local Hash were found to be the best performing FO protocols by \cite{wang2017locally}, however while the former benefits from the privacy amplification effect the latter does not. Wang et al. therefore propose to adaptively select the best FO protocol based on the size of the item domain $|\mathcal{I}|$, $\epsilon$ and $\ell$. Due to the amplification effect Generalized Random Response should be used for any
$$
|\mathcal{I}| > (4 {\ell}^2 - \ell) \cdot e^{\epsilon} + 1
$$
They further note that the choice of $\ell$ is crucial a choice too small can lead to estimation errors while a choice too big may magnify noise in the estimation of the frequency oracle. 

\begin{figure}[!hbt]
    \centering
    \includegraphics[width=0.75\textwidth]{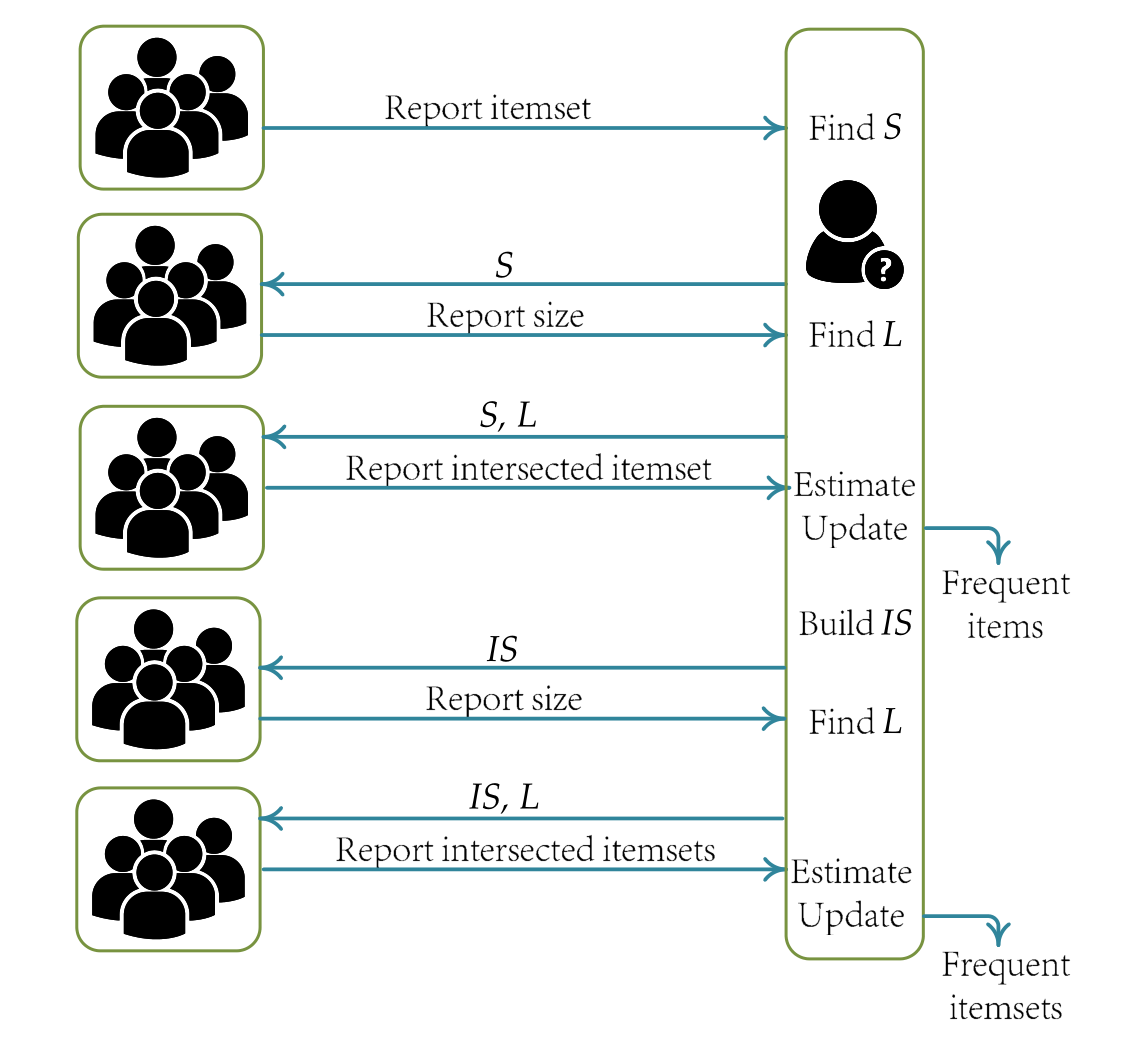}
    \caption{Illustration of the SVIM and SVSM protocols from \cite{wang2018locally}. Users (left) are partitioned into five groups. Aggregator (right) runs SVIM to identify frequent items with first three groups, then SVSM to identify frequent itemsets with last two groups.}
\end{figure}

The SVIM protocol works similarly to LDPMiner but makes use of the privacy amplification effect to achieve a higher accuracy. The protocol works in four steps. The users are partitioned in three groups, each participating in one tasks. It is shown in~\cite{wang2017locally} that this provides improved accuracy. In the first step users report inputs with small $\ell$ and the aggregator identifies heavy hitter candidates which are subsequently broadcasted back to the users. Second, using a standard FO protocol, users report back to the aggregator the number of candidate items they have. The aggregator can then choose an appropriate $\ell$ which is broadcasted to users. Third, users can report their candidate items using PSFO with the given $\ell$. This allows the aggregator to create frequency estimates for these items. Finally, the aggregator identifies the $k$ most frequent items. The known size distribution from step two can be used to further correct undercounts of these items.

The task of mining itemsets is much more challenging as there are exponentially more candidates to consider. To address this Wang et al. introduce SVSM which can find frequent itemsets effectively. The protocol first applies SVIM to find frequent items and to reduce the range of possible itemsets to a manageable size. Using the list of $k$ most frequent items a candidate set of itemsets $IS$ can be constructed. One can then map this problem to the regular frequent item mining problem and apply SVIM to find the most frequent itemsets.

SVIM significantly outperforms LDPMiner while SVSM solves a previously open problem. Both are the state-of-the-art in frequency estimation on set-valued inputs.

\subsection{Private Spatial Data Collection}

Many services such as \emph{Google Maps} or \emph{Waze} benefit from collection of user data to identify popular locations and to create traffic congestion maps. Given a large number of users' and their location data, we would like to learn their distribution over a spatial domain while maintaining user privacy. This problem has previously been studied in the centralized differential privacy setting \cite{cormode2012differentially,qardaji2013differentially}. Chen et al. \cite{chen2016private} introduce the notion of \emph{personalized local differential privacy} (PLDP) and propose a framework that can learn the user distribution over a spatial domain (location universe) while guaranteeing PLDP for each user.

As the location universe is a large domain in most real-world settings, $\epsilon$-LDP can not be maintained when one wants to obtain reasonably accurate results. Chen et al. therefore introduce the notion of \emph{personalized local differential privacy} (PLDP) which enables users to control their own privacy settings individually by setting their own privacy parameters:
\begin{definition}
Given the personalized privacy specification $(\tau,\epsilon)$ of a user $u$, a randomized algorithm $\pi$ satisfies $(\tau,\epsilon)$-personalized local differential privacy for $u$, if for two locations $l, l' \in \tau$ and any $O \subseteq Range(A)$,
$$
\frac{\text{Pr}[\pi(l) \in O]}{\text{Pr}[\pi(l') \in O]} \leq e^{\epsilon}
$$
where the probability space is over the coin flips of $\pi$.
\end{definition}
Here, $\tau$ determines a user's \emph{safe region} which they feel do not mind to reveal but don't want others learning about any more fine-grained location data than that. PLDP is a generalized version of $\epsilon$-LDP as $\tau = \mathcal{L}$ for a location universe $\mathcal{L}$ implies regular $\epsilon$-LDP.

Chen et al. then introduce the private spatial data aggregation (PSDA) framework. To estimate the number of users in a given region they define the personalized count estimation protocol (PCEP). Given $n$ users' locations, their privacy specifications and a confidence parameter $\beta$ which determines the accuracy, this protocol estimates user counts for each region. The reported locations for each user are perturbed using a local randomizer which guarantees $(\tau,\epsilon)$-PLDP. Next, they define user groups as sets of users with the same safe region and partition these user groups into clusters to minimize the maximum absolute error $\max_{l \in \mathcal{L}} |\hat{s_l} - s_l|$ for true and estimated user counts $\hat{s_l}$ and $s_l$ in the location universe $\mathcal{L}$. The untrusted server can then apply PCEP for each cluster $C_i \in \mathcal{C}$ with a confidence parameter $\frac{\beta}{|\mathcal{C}|}$ so that the overall confidence level is guaranteed to be $\beta$. Finally the server calculates counts for all locations by combining the estimates from all clusters accordingly. 

\section{Open problems in LDP}

As Local Differential Privacy is still a relatively new field some questions still remain open. Recent work by Avent et al.~\cite{avent2017blender} has proposed a new hybrid model which allows users to choose between the local and the centralized model. Since accuracy can suffer in the pure LDP setting, as LDP gives users very strong privacy guarantees, this hybrid approach can provide improvements in accuracy. The approach by Avent et al. provides a \emph{blended} algorithm for local search that can combine information from both the central and the local model. As their work focuses on local search, which is an application of heavy hitter identification, it remains open how this hybrid model can be applied to other problems in LDP and how much of a benefit it can provide.

\medskip

It has been previously demonstrated that deep learning models can be trained while preserving differential privacy. It has also been demonstrated that some machine learning algorithms like linear regression, logistic regression and SVM classification can be successfully applied in the LDP setting \cite{nguyen2016collecting}. It is still open however whether deep learning models can be trained in the LDP setting.

\medskip

Recent research by Qin et al.~\cite{qin2017generating} has presented a new approach to generate representative synthetic social graphs in an LDP setting using multiple rounds of interactions between the server and the user where based new queries are made based on previous responses. Qin et al. were able to improve the protocol's accuracy using this multi-phase interaction. It is still open how other LDP protocols can benefit from multiple interactions.

\section{Conclusion}

In recent years LDP has risen to popularity and seen its first real-world deployments. We have presented and discussed some of these deployments such as \cite{erlingsson2014rappor} and \cite{apple2017ldp} that are used by hundreds of millions of users every day. Today LDP is the state-of-the-art approach to give strong privacy guarantees to users while enabling organizations to collect usage statistics on their products using locally private protocols. We have given an overview over the guarantees $\epsilon$-LDP can provide as well as the pure LDP setting by \cite{wang2017locally}. As frequency estimation is at the core of many LDP protocols we have given an in-depth overview over different and optimized frequency oracle protocols in the pure LDP setting introduced by \cite{wang2017locally}. Furthermore, we have presented the core problems in LDP such as locally private heavy hitter identification, itemset mining and spatial data collection. Finally, we have given a short outlook on the future of LDP and recent research directions. Further research is necessary to determine how hybrid models~\cite{avent2017blender}, that give the user the freedom to choose between the central and local setting, and multi-phase interaction protocols~\cite{qin2017generating} can be applied to other LDP problems and what their theoretical boundaries are.

%
%
%
%
\bibliographystyle{splncs04}
\bibliography{bibliography.bib}
\end{document}